# Optimization of portfolios with cryptocurrencies: Markowitz and GARCH-Copula model approach

Vahidin Jeleskovic,[a] Claudio Latini,[b] Zahid I. Younas,[c] Mamdouh A. S. Al-Faryan,[d]

**Abstract:**

The growing interest in cryptocurrencies has drawn the attention of the financial world to this innovative medium of exchange. This study aims to explore the impact of cryptocurrencies on portfolio performance. We conduct our analysis retrospectively, assessing the performance achieved within a specific time frame by three distinct portfolios: one consisting solely of equities, bonds, and commodities; another composed exclusively of cryptocurrencies; and a third, which combines both 'traditional' assets and the best-performing cryptocurrency from the second portfolio.To achieve this, we employ the classic variance-covariance approach, utilizing the GARCH-Copula and GARCH-Vine Copula methods to calculate the risk structure. The optimal asset weights within the optimized portfolios are determined through the Markowitz optimization problem. Our analysis predominantly reveals that the portfolio comprising both cryptocurrency and traditional assets exhibits a higher Sharpe ratio from a retrospective viewpoint and demonstrates more stable performances from a prospective perspective. We also provide an explanation for our choice of portfolio optimization based on the Markowitz approach rather than CVaR and ES.

[a] Humboldt-Universität of Berlin, email: vahidin.jeleskovic@hu-berlin.de
[b] Independent researcher, email: claudio.lat95@gmail.com
[c] Berlin School of Business and Innovation, email: zahid1132_gcu@yahoo.com
[d] University of Portsmouth, email: al-faryan@hotmail.com

# 1. Introduction

Introduction of crypto currencies in financial system becomes the hot issue of investment among the worldwide investors. Like traditional currencies the cryptocurrencies also offer the holder of currency the purchase of goods and services. However, contrary to legal tender payment methods, the cryptocurrency-system is based exclusively on the trust of the parties in cryptocurrencies and crypto tokens sine there is no law yet that obliges to accept cryptocurrencies as a means of payment (Gogo, 2019). A public register of filing of the exchanges *(called block chain)* to which everyone can access, is the base of the fiduciary system; all the operations occurring between the holders of cryptocurrencies are therefore validated by a *miner*,[2] whose task is to guarantee and encrypt the transactions that have taken place by entering them in the public register (Tuwiner, 2019).

Despite the absence of regulatory requirements, the acceptability of cryptocurrency as a medium of exchange got the tremendous attention among investors and its volume of trade has also gone up by leaps and bounds in a very small period of time.[3] The reason behind this success is the financial crisis of 2008 that not only resulted in the loss of hard-earned investments of the people across the globe, but also forced them to look for protection against such crisis and alternatives for safer investment (Pinudom et al., 2018). Further, cryptocurrencies may be unrelated to the prevailing economic situation of any country, and as an alternative investment also provide the opportunity for financial diversification (Bodie, Kane, and Markus, 2014; Trimborn, Mingyang, and Härdle, 2017; Krueckeberg and Scholz, 2018).[4] Moreover, adding the cryptocurrency to the portfolio also benefit the investor against the various monetary policy related risks attached with traditional currency and thus finally improve the performance of portfolio (Anyfantaki and

---

[2] Nowadays, this kind of validation depends on the type of consensus.
[3] The three cryptocurrencies with the highest market capitalization (Bitcoin, Etherium and XRP) on 31st December 2018 reached a trade volume equal to 579.582.414 units for an amount of Euro equal to 113.336.458.098,33. Data provided by https://coinmetrics.io.
[4] According to the Modern Portfolio Theory and the principle of diversification, the asset, which shows a low level of correlation with other assets, is considered a good alternative investment (Elton et al., 2009).

Topaloglou, 2018; Mayer, 2018; Elendner et al., 2017; Briere, Oosterlinck, and Szafarz, 2015). Financial diversification, in addition to being a risk-reducing instrument, is also a motive of achieving greater profit. The investors with such motives are called "risk-seekers", i.e. those who manage to make a profit from the high volatility of the markets.

Since investor has started opting the cryptocurrency as an integral part of their portfolio and capital allocation, thus here a question arises that, whether cryptocurrencies result in the more beneficial optimization of their portfolio or not? Portfolio optimization can be defined as the maximum benefit that can derive from an allocation of financial resources in relation to the risk-return profile of the investor. Markowitz (1952), used average, variance, co-variance and Pearson's linear correlation for construction of optimal portfolio. However, construction of Markowitz theory of optimal portfolio selection does not consider that distribution of historical returns of financial assets may not be Gaussian. While, his point of view is widely rejected in financial literature with empirical evidence that financial historical series are very often characterized by phenomena of asymmetry and leptokurtosis (Sheikh and Qiao, 2009) or even more dramatic stylized facts (Cont, 2001). Assuming that financial returns are normally distributed lead to their underestimation and inaccurate quantification of risk (Pinudom et al., 2018). Similarly, under the Markowitz portfolio selection, measuring risk only by variance of historical returns is very general for specific asset and also incorrect. Moreover, Markowitz used Pearson´s linear correlation as a tool to measure the association among assets for construction of portfolio, however, the Pearson's correlation ignores the non-linear association among the assets and thus conceal their dependency structure (Rachev, Sun and Stein, 2009). To overcome above mentioned issues conditional autoregressive and generalized heteroscedastic models (GARCH) are also widely used to predict the future evolution of variance in research related to virtual currencies (Saha, 2018; Chu et al., 2017; Katsiampa, 2017; Chen et al., 2016; Ardia et al. 2018b). While, the study of Zekokh (2019) has extended the analysis of Ardia et al. (2018b) on Bitcoin, Ethereum, Ripple and Litecoin, he considered a Model Confidence Set (MCS) procedure to select the GARCH models for the Markov regime switching for each cryptocurrency.

Further, the linear correlation coefficient may also be inadequate in identifying the structure of interdependence of the assets (Embrechts et al., 1999; Szegö, 2005). Thus related literature suggest that financial association among the financial assets should be determined by Copula models as it facilitates to model for each asset, the marginal probability distribution and to evaluate separately, and successively, the multivariate dependence of all assets in a portfolio (Sklar, 1959; Cherubini, Luciano, and Vecchiato, 2004; Nelsen, 2007). The Copulas therefore focus on determining a correlation structure between the univariate distributions of the assets and are more flexible than the standard multivariate distributions (Kakouris and Rustem, 2014). This in turn makes possible to break the link with the Gaussian distribution (Lujie Sun and Manying Bai, 2007; Manying Bai and Lujie Sun, 2007).

As GARCH-Copula models provide the better forecast some mutual properties future returns than bootstrapping methods, they are frequently applied in the field of portfolio diversification (Kresta, 2015; Bouoiyour et al., 2017; Osterrieder et al., 2016).[5] However, the Copula models are often applied only in bivariate contexts due to the fact that the multivariate Copula has a kind of complexity that increases as the size increases, tending to lose precision of fitting on the tails (Deng et al., 2011). Considerable efforts have been made to increase the flexibility of the multivariate Copula models. The Vine-Copulas models are the result of this endeavour and it was further modified into its subclasses of C-Vine and D-Vine (Joe, 1996; Bedford and Cooke, 2001). Later on many studies have not only used these Copulas and their subclasses into their research on financial returns, portfolio management, exchange rate management, but also proved empirically that Vine-Copula approach outperform the multivariate t-Copula, especially when returns have asymmetry and a different dependency structure between pairs of financial assets (Aas et al., 2009; Schirmacher and Schirmacher, 2008; Mendes et al., 2010; Fischer et al. 2007; Saha, 2018).

In the light of relationship of risk and return, this study aims to explore the impact on risk and return structure of portfolio if cryptocurrencies as financial assets are considered as a part of optimized portfolios. The study uses GARCH models to quantify the variance of returns whereas among Copula Models Vine-Copula models are applied to capture the correlation structure between the different assets. The use of a GARCH process combined with a Copula model allows to divide the risk due to several factors (standard deviation, skewness and kurtosis and co-movements between assets), in different mathematical steps trying to specify the risk of portfolio in the best possible way. In this paper, the analysis is based on a comparative perspective. Three portfolios are evaluated on the basis of the performance achieved: the first one composed only by traditional assets, we call it "Traditional portfolio", the second one composed by the same assets with the inclusion of a cryptocurrency, thus "Traditional Crypto portfolio", and the third one makes up by only cryptocurrencies - "Cryptos portfolio". The three portfolios are constructed under consideration of three different specifications regarding returns' distribution: the first one assumes a Normal distribution of the historical returns; the second one and third one combine both multivariate Copula model and Vine Copula model, respectively, with the GARCH modeling of the variance. Our results show that combining the traditional assets with the crypto-assets may lead to the higher gains in terms of reward-risk ratio, and thus should be considered in the more intensive future work. The final contribution of the study is also to evaluate that, which of the model among proposed models provide the better performance and stability of portfolio over time considering different optimality criteria.

---

[5] See for examples: Krzemienowski and Szymczyk, (2016); Paolella and Polak, (2018); Shekhar and Trede, (2017).

In this context, we wish to emphasize our deliberate choice of the Markowitz approach over methods based on Conditional Value at Risk - CVaR and its counterpart Expected Shortfall – ES (see, for example, Ziemba and Vickson, 2006; Rockafellar and Uryasev, 2002; Krokhmal et al., 2002; and Stoyanov et al., 2007). There are several compelling reasons behind this decision. The CVaR method takes into account the so-called fat tails in the return distribution of the underlying asset, effectively measuring the risk of significant losses when optimizing portfolios. However, it's important to note that returns in regulated markets stem primarily from the fundamental value and the trading process of the asset. In contrast, the prices of cyberassets can be heavily influenced by a multitude of negative exogenous events that are not directly linked to their fundamental values or trading processes. These events can include cyber-attacks and their media coverage, the impact of influential figures on social media (such as Elon Musk's effect on Dogecoin), failures in implementing mainnet and significant protocol upgrades, and more. Researchers often overlook this distinction and treat these events as if they were endogenous, which can lead to misleading investment strategies for cryptoassets. To summarize, while extreme events in traditional financial markets are rare but systematic and tend to recur, the world of crypto markets witnesses a vast number of extreme events that are often unique and do not repeat. Furthermore, within the crypto investing community, a common mantra prevails: "trend is your friend," and investors tend to follow the principle of "buying on the dip and selling on the peak." This implies that investors are more focused on identifying and capitalizing on positive trends, where the middle of the return distribution holds greater importance than isolated extreme events at the tails. Consequently, our choice of the Markowitz approach aligns with the needs of medium and long-term investors rather than those with a very short-term perspective.

In conclusion, our decision has been to utilize data predating the emergence of the COVID-19 pandemic and the Ukraine conflict. These events have given rise to recent economic and energy crises, as well as exceptional market conditions. Our rationale is rooted in the belief that using data from periods prior to these crises provides a more faithful representation of the ordinary and stable economic environment, which we deem to be the prevailing state for the majority of the time.

The remaining sections of the paper are structured as follows. In Section 2, we introduce Copula models and their relevance to our analysis. Section 3 provides a detailed explanation of the methodology employed for data analysis and optimization. Moving forward to Section 4 we present the examined data and report the results obtained from each portfolio. Section 5 is dedicated to a comprehensive discussion of the main findings, conclusions, and an outlook for future research and implications.

## 2. Methodology

### 2.1. Location Scale Model

The Location Scale Model (or Location Scale Family) is the theoretical basis on which the entire mathematical system of the GARCH-Copula model is based. A random variable $X$ is said to belong to the location–scale family when its cumulative distribution (CDF) is a function of $\frac{x-a}{b}$:

$$F_X(x \mid a, b) = F\left(\frac{x-a}{b}\right); a \in R, b > 0; \qquad (1)$$

The two parameters $a, b,$ are respectively the location and the scale parameters.[6] The variable

---

[6] "The location parameter $a$ is responsible for the distribution's position on the abscissa. An enlargement (reduction) of $a$ causes a movement of the distribution to the right (left). The location parameter is either a measure of central tendency of a distribution. The parameter $b, b > 0$, is the scale parameter. It is responsible for the dispersion or variation of the variate X. Increasing (decreasing) $b$ results in an enlargement (reduction) of the spread and a corresponding reduction (enlargement) of the density" (Rinne, 2010).

$$Z := \frac{X - a}{b} \qquad (2)$$

is called the reduced or standardized variable, where $z$ is a realization of $Z$. The reduced variable $Z$ has $a = 0$ and $b = 1$. If $Y$ has a cumulative distribution function $F_Z(z) = P(Z \leq z)$, then $X = a + bZ$ has a cumulative distribution function $F_X(x) = F_z\left(\frac{x-a}{b}\right)$. In other worlds, "For any random variable $Z$ whose CDF belongs to such a family, the CDF of $X = a + bZ$ also belongs to the so-called location–scale family. The use of this model in finance can certainly be traced back to the studies of Azzalini, (1985), with the introduction of a third parameter $c$, that is a parameter of "shape".

**Lemma 1:** If $f_0$ is a one-dimensional probability density function symmetric about 0, and $G$ is a one-dimensional distribution function such that $G'$ exists and is a density symmetric about 0, then

$$f(z) = 2f_0(z)G\{w(z)\} \, (-\infty < z < \infty) \qquad (3)$$

is a density function for any odd function $w(\cdot)$.[7]

On the basis of the Lemma 1 and the respective demonstration, assuming $f_0 = \phi$, $G = \Phi$ (respectively the PDF, and the CDF of the Normal Standard distribution) and $w(z) = c \cdot z$ ($c$ is a constant). Azzalini (1985) was able to affirm that: if Z is Skewed Normal distributed ($Z \sim SN(c)$) and $X = a + bZ$, where $a \in R, b > 0$, then it is possible state that $X \sim SN(a, b^2, c)$.[8] Thanks to this contribute, other authors have been able to give important reflections: Arellano-Valle et al. (2004) extended the number of parameters through which it is possible to represent the distribution that remains from a process of "standardization": the location-scale skew-generalized normal distribution (SGN) is defined as that $X = a + bZ$, where $Z \sim SGN(c, d)$ and $X \sim SGN(a, b^2, c, d)$. Only in more recent studies Kumar et al. (2018) have demonstrated the possibility of representing a reduced variable through a 3-factor parameterization of the shape, keeping the connection with the Location-Scale model.

---

[7] See Azzalini (1985).
[8] Azzalini (1985) formalizes also the case of Skewed t – distribution.

## 2.2 Autoregressive models

A necessary but not sufficient condition is to accept the assumption that the studies underlying the Location-Scale model are valid, considering that returns are the variable under the microscope for this analysis. Empirical evidence (beginning with Engle (1982)) shows that it is possible to consider the variance variant over time. This logic assumption, combined with the Location-Scale model, allows to describe the evolutionary process of returns as:

$$R_t = a_t + b_t Z_t \tag{4}$$

Where $a_t, b_t^2, Z_t$ are respectively the conditional mean, conditional variance and the innovation of $R_t$ process. This type of formulation allows therefore to model the Location process, as an autoregressive moving average process (ARMA(m,n) process):

$$\boldsymbol{a_t = a + \sum_{j=1}^{m} \phi_j (R_{t-j} - a)} + \sum_{j=1}^{n} \boldsymbol{\theta_j (R_{t-j} - a_{t-j})} \tag{5}$$

and parameter of Scale process, using generalized autoregressive conditional heteroscedastic process (GARCH(q,p) models):

$$b_t^2 = \omega + \sum_{j=1}^{q} \varphi_j (R_{t-j} - a_{t-j})^2 + \sum_{j=1}^{p} \psi_j b_{t-j}^2 \tag{6}$$

The Equation (7) shows the formulation provided by Bollerslev (1986), for the first GARCH model (simple GARCH). After this first model the research has proposed until today many GARCH models capable in different ways of capturing effects of asymmetry and intensity of the conditional variance.

The integrated GARCH model (Engle and Bollerslev, 1986), denoted by IGARCH(p ,q), is a particular case of the simple GARCH, because $\sum_{j=1}^{q} \phi_i + \sum_{j=1}^{p} \beta_i = 1$. The condition makes the model strictly stationary.

The exponential GARCH model (Nelson, 1991), denoted by EGARCH (q, p), has:

$$\ln(b_t^2) = \omega + \sum_{j=1}^{q} \left( \varphi_j Z_{t-j} + \gamma_j (|Z_{t-j}| - E|Z_{t-j}|) \right) \\ + \sum_{j=1}^{p} \psi_j \ln(b_{t-j}^2) \qquad (7)$$

for $\varphi_j > 0, \psi_j > 0, \gamma_j > 0$ and $\omega > 0$. $\varphi_j$ captures the sign effect, and $\gamma_j$ captures the size effect of the previous standardized innovation. The persistence parameter for this model is $\psi_j$.

The GJRGARCH (q, p) model due to Glosten et al. (1993) has:

$$\sigma_t^2 = \omega + \sum_{j=1}^{q} \left( \varphi_j (R_{t-j} - a_{t-j})^2 + \gamma_j I_{t-j} (R_{t-j} - a_{t-j})^2 \right) \\ + \sum_{j=1}^{p} \psi_j \sigma_{t-j}^2 \qquad (8)$$

for $\varphi_j > 0, \psi_j > 0, \gamma_j > 0$ and $\omega > 0$, where $I_{t-j} = 1$ if $(R_{t-j} - a_{t-j}) \leq 0$ and $I_{t-j} = 0$ if $(R_{t-j} - a_{t-j}) > 0$. $\gamma_j$ represents an asymmetry parameter. A positive shock will increase volatility by $\varphi_j$; a negative shock will increase volatility by $\varphi_j + \gamma_j$.

Simple GARCH, IGARCH, GJRGARCH and EGARCH combined with the ARMA process are used to fit the conditional volatility and conational mean of the financial assets. For the analysis are considered a maximum order of one for the ARMA part, and a maximum of order three for the GARCH models. The conditional distribution of the innovation is supposed be Normal or Skewed Normal. So, an iterative procedure estimates, for all 4 ARMA-GARCH processes, the best model for each combination of lags until the maximum settled, replete this procedure 2 times, one for each distribution hypothesis of the innovation.[9] The lower BIC value is chosen as criterium to identify the most adapt model.

## 2.3 Copula and Vine-Copula models

---

[9] The algorithm estimates 4 different types of GARCH, each of which has 4 different dispositions of the parameters of ARMA process and 9 possible dispositions of the parameters of GARCH process. All this is done for each of the assumed conditional distributions of the innovations, for a total of 288 models for each financial asset.

Copulas are models through which it is possible to isolate the dependence structure of a multivariate distribution (Nelsen, 2007). Let $H$ be a n-dimensional distribution function with marginal distribution functions $F_j(Z_j; \theta)$. Then there exists a copula $C$ such that:

$$H(Z_1, \ldots, Z_n) = C(F_1(Z_1; \theta), \ldots, F_n(Z_n; \theta); \delta) \qquad (9)$$

Where $C(F_1(x_1; \theta), \ldots, F_n(x_n; \theta); \delta)$ is the Copula associated to H and $\delta$ is the vector of dependence parameters of $C$. The Copula is unique if the marginals $F_j(Z_j; \theta)$ are continuos (Sklar, 1959). The unicity of the Copula was demonstrated by Sklar (1959). A large number of Copulas have been proposed in the literature, and each of these imposes a different dependence structure on the data (Trivedi and Zimmer, 2007). Joe and Xu (1996), Trivedi and Zimmer (2007), Balakrishnan and Lai (2009) and Nelsen (2007) provide a detailed overview about the properties of Copulas (Elaal, 2017). Gaussian (Normal) copula, Student's copula, Gumbal and Clayton copula are taken into consideration for this analysis. The Normal copula, proposed by Lee (1983), takes the form:

$$H(Z_1, \ldots, Z_n) = \Phi_G(\Phi^{-1}(F_1(Z_1; \theta)), \ldots, \Phi^{-n}(F_n(Z_n; \theta)); \delta) \qquad (10)$$

where $\Phi$ is the CDF of the standard Normal distribution, and $\Phi_G$ is the standard multivariate Normal distribution with correlation parameter restricted to the interval $(-1, 1)$. A Copula with two dependence parameters is the multivariate t-distribution with $v$ degrees of freedom and correlation $\delta$,

$$H(Z_1, \ldots, Z_n) = t(t_{m_1}^{-1}(F_1(Z_1; \theta)), \ldots, t_{m_2}^{-n}(F_n(Z_n; \theta)); \delta, v) \qquad (11)$$

where $t_m^{-1}$ denotes the inverse of the CDF of the standard univariate t-distribution with $m$ degrees of freedom. The parameter $v$ controls the heaviness of the tails. The Clayton (1978) copula, proposed by by Kimeldorf and Sampson (1975), takes the form:

$$H(Z_1, \ldots, Z_n) = \left(F_1(Z_1; \theta)^{-\delta} + \cdots + F_n(Z_n; \theta)^{-\delta} - 1\right)^{-1/\delta} \qquad (12)$$

with the dependence parameter $\delta$ restricted on the region $(0, \infty)$. As $\delta$ approaches zero, the marginal become independent. The Clayton copula cannot account for negative dependence. It has been used to study correlated risks because it exhibits strong left tail dependence and relatively weak right tail dependence (Trivedi and Zimmer, 2007).

The Gumbel copula (1960) takes the form:

$$H(Z_1, \ldots, Z_n) = \exp\left(-\left(-\log F_1(Z_1; \theta)^\delta + \cdots \right.\right.$$
$$\left.\left. - \log F_n(Z_n; \theta)^\delta\right)^{1/\delta}\right) \tag{13}$$

The dependence parameter is restricted to the interval $[1, \infty)$. Values of 1 and $\infty$ correspond to independence. Similar to the Clayton copula, Gumbel does not allow negative dependence, but it contrast to Clayton, Gumbel exhibits strong right tail dependence and relatively weak left tail dependence (Trivedi and Zimmer, 2007).

The Student's copula and the Gaussian copula belong to the family of Elliptical copulas, the Gumbel and the Clayton copulas instead are part of the family of the Archimedeans. The main difference between two families of Copulas is that the Elliptical ones have the possibility to specify the level of correlation between the marginal distributions, contrary to the other. The Archimedean ones have the merit of modelling well the extreme values of the multivariate distributions. These characteristics, therefore, make one more or less suitable than the other depending on the case. In a portfolio allocation problem with more than two assets, it is improbable that a single correlation parameter (as in the case of Elliptical copulas) can represent a correlation structure of n-dimensions.

The possibility of using the elliptical functions even in a context that concerns more than two dimensions, making the Copula approach even more flexible, led Joe in the 1995 to elaborate the first Vine Copula model. A Vine Copula is a factorization of multivariate Copula densities into (conditional) bivariate Copula densities. For example:

Let $c(F_1(Z_1; \theta), F_2(Z_2; \theta), F_3(Z_3; \theta); \delta)$ be the density function of a 3-dimentional Copula, the respective Vine Copula structure is composed by three parts:

$$conditional\ density\ pair = c_{hg|m}\left(F_{h|m}(Z_h|Z_m); F_{m|g}(Z_m|Z_g)\right) \tag{14}$$

$$uncoditional\ density\ pairs$$
$$= c_{mg}\left(F_m(Z_m); F_g(Z_g)\right) \cdot c_{hg}\left(F_h(Z_h); F_g(Z_g)\right) \tag{15}$$

$$maginal\ density\ functions = f_g(Z_g) \cdot f_g(Z_g) \cdot f_h(Z_h) \tag{16}$$

$$c(F_1(Z_1; \theta), F_2(Z_2; \theta), F_3(Z_3; \theta); \delta)$$
$$= (c.d.pair) \cdot (u.d.pairs) \cdot (m.d.functions) \tag{17}$$

As it possible to see, the adaptability of Copula models makes them much more complex than multivariate Copulas in terms of estimation. In fact, these models are composed by conditional copula densities, joint copula densities and the marginal probability functions. The initial sequence of these assets is not known a priori: this means that it is not possible to know how to couple the financial assets from time to time, in order to form the various Copula bivariate. Moreover, the determination of the initial sequence is very often subordinate to the choice of the structure of the Vine Copula. In literature there are three main structures studied: The Regular-vine copula, the Canonical-vine copula and the Draw able-vine copula; the last two are a subcategory of the first one. The C-vine has a star structure where a unique node is connected to all other ones. The D-vine has path structure where each asset has a single link with the next one. The R-vine is built considering the $\frac{n!}{2} \cdot 2^{\binom{n+1}{2}}$ possible combinations between the assets (Morales-Nàpoles, 2010). Regardless of the chosen structure, it should be noted that the link between each pair of assets is unique; this allows to consider as far as the C- and D- copulas $\frac{n!}{2}$ possible dispositions are concerned. Moreover, while for these last two there is a rule that establishes how to construct the pairs following to the starting tree, the structure of the R-vines is always the result of a computational procedure. Once an initial order of the marginal distributions has been determined and a methodology able to proceed in the construction of the pairs has been identified, it is necessary to identify the type of Copula that binds each pair, estimating its relative parameters.

The two different methodologies (Vine and multivariate) are comparable using the BIC and AIC criteria. The goodness of fitting (GoF) of the both Copula models structure will be measured through the test proposed by Genest et al. (2009). This functional and certainly not exhaustive excursus on Copula models can be concluded with the following formal definition:

**Definition 1.** A d-dimensional Copula, $C: [0; 1]^d :\to [0; 1]$ is a cumulative distribution function (CDF) with uniform marginal.

It is precisely this last definition that creates the link between the Location-Scale model, the GARCH and Copula models. The definition indicates that the only values that can be used as input to create a Copula model are between 0 and 1. This means that the standardized residues, obtained from the GARCH model, need to be converted into probability, through the identification of a distributive function. In fact, according to the integral transformation of probability, it is possible to transform random variables belonging to any distribution into uniformly distributed random variables; the statement is valid provided that the distribution used as a means of transformation is the real one. In order to identify the true distribution, the Kolmogorov-Smirnov test (K-S test)

(Massey, 1951) and the Uniformity test (PIT test) (Diebold, Gunther Tay, 1997) have been used on a sample of 5 different distributions (Standard normal, t-Student, Skewed t-Student (Skwd. t-Stud.), Generalized Error distribution (GED), Skewed Generalized Error distribution (Skwd. GED).

## 2.4 Portfolio optimization

A key step towards a quantitative approach to the issue of strategic asset allocation was taken by Harry Markowitz, in his article "Portfolio Selection" published on the Journal of Finance in 1952. Its model assumes essentially that the investors are rational and are interested in maximizing the returns and minimizing the risk, furthermore they have free access to correct information on the returns and risk; the markets are efficient and absorb the information quickly and perfectly. Under these hypothesises, Markowitz consider the risk how the oscillation of the returns around their average. In these terms we can consider the variance of the returns of a portfolio as its risk and the average of these returns as the expected return of the portfolio. In statistical terms it is therefore necessary to minimize the variance (or standard deviation) and maximize the average, adjusting the weights that each asset has inside of the portfolio. Formally:

$$\min_{w_i} (\sigma_p^2) = \min_{w_i; w_j} \left( \sum_i \sum_j w_i w_j \sigma_{ij} \right) \quad (18)$$

$$E(Rp) = \sum_{i=1}^{n} w_i R_i \quad and \quad 0 \le w_i \le 1 \quad (19)$$

This model is not free of limits: first of all, this way of optimising a portfolio limits the number of assets to be chosen, contrary to the benefit of diversification, creating violent and drastic shifts in the composition of the portfolio over time; moreover, the past performance of a portfolio has little predictive power of its future performance (Braga, 2016). In the empirical work, to determine the weight of each asset in portfolio, the maximization of Sharpe ratio[10] is set as target. In literature this is known how Tangency portfolio[11].

(20)
$$\max_{w} \frac{E(Rp)}{\sigma_p}$$
$$\text{Subject to} \sum_{i} w_i = 1$$
$$0 \leq w_i \leq 1$$

The use of this model together with the above limits allows a transversal analysis of the problem: on the one hand it allows to assess whether a portfolio that contains cryptocurrencies is actually more performing than those composed only of traditional assets, on the other hand it allows to see if a Location-Scale-GARCH-Copula model approach allows to solve (at least partially) the problems mentioned above. Finally, the portfolio's performance is assessed using the Sharpe index in the version proposed by (Pezier and White, 2006)[12].

## 3. Data and results

The first step towards a comparative analysis of the three different portfolios begins with the choice of assets. This first phase is conducted by taking into account the past performance of the returns of a set of assets over a period ranging from 01 January 2017 to 31 December 2018.[13] We point out the data used in this paper is from a period before the corona pandemic and the war in Ukraine. We believe that the data before the recent economic and energy crises, caused by the corona pandemic and this war respectively, is more representative for the normal state of economic environment. The first portfolio, called "Traditional portfolio" (or Trad), is composed exclusively of assets belonging to the investment universe that is represented by equities, government bonds and commodities[14]. As for the stock market, the *Dow Jones* and *Eurostoxx 50* indices provided a basis for stock picking. In fact, out of the total 80 equity assets, the three most preforming ones for both indices were chosen: Microsoft (MSFT), Boing (BA) and Visa (V) from Dow Jones, RWE AG (RWEG), Amadeus (AMA) and Kering S.A. (PRTP) form Eurostoxx 50. Comparing the government bond yields of France, Germany, Italy, UK and USA at 10 years, it was found that US government bonds were the best in terms of risk-return. In order to achieve the most accurate possible replication of this type of return, it was decided to include the ETF (Exchange Traded Fund) IUSM in the portfolio. Finally, for the choice of the 4 commodities, the performance analysis was carried out, taking into account the most liquid markets. The most performing commodities were Palladium (PA), Copper (HG), Gold (GC). Crude oil (CL) is forcibly added to guarantee the most diversification possible. The second portfolio, called "Cryptos portfolio" (or Cryptos), is composed by 10 cryptocurrencies, which were selected from the 200 with the highest market capitalization at 31 December 2018; the cryptocurrencies that performed best were PIVX (PIVX), Ethereum (ETH), XRP (XRP), Stellar (XLM), NEO (NEO), Decred (DCR), Waves (WAVES), Verge (XVG), Unobtanium (UNO) and Groestlcoin (GRS).[15] The third portfolio, called "Traditional Crypto portfolio" (or Trad+Cry) is composed of all the assets of the "Traditional portfolio" with the addition of the most performing cryptocurrency among those selected for the "Cryptos portfolio".

---

[10] The Sharpe ratio considered is the original one (Sharpe, 1994).
[11] The Tangency portfolio is identified at the point where the Capital Market Line tinges the (Markowitz) efficient frontier.
[12] This version (Adjusted Sharpe Ratio) allows to assess the risk of a portfolio not only in terms of standard deviation, but also in terms of kurtosis and skewness.
[13] The performance analysis is carried out by calculating the Sharpe index in the version proposed by Pezier and White (Pezier & White, 2006); it is also assumed the absence of exchange rate risk.

**Table 1: Performances of cryptocurrencies and traditional financial assets from 1 January 2017 to 31 December 2018.**[16]

| Asset name | PRTP | RWEG | AMA | IUSM | BA | Va | MSFT | PA | HG | GC | CL |
|---|---|---|---|---|---|---|---|---|---|---|---|
| Expected return | 0,14% | 0,10% | 0,06% | -0,02% | 0,12% | 0,09% | 0,08% | 0,10% | 0,02% | 0,01% | 0,00% |
| Standard Deviation | 0,017 | 0,018 | 0,012 | 0,005 | 0,016 | 0,013 | 0,014 | 0,015 | 0,013 | 0,010 | 0,017 |
| Kurtosis | 0,073 | -0,910 | -0,408 | -0,026 | 0,023 | -0,170 | 0,129 | -0,438 | -0,049 | 0,222 | -0,851 |
| Skewness | 5,945 | 9,308 | 1,414 | 0,765 | 3,220 | 3,094 | 3,476 | 3,033 | 1,564 | 5,412 | 2,528 |
| Adjusted Sharpe | 0,080 | 0,054 | 0,051 | -0,042 | 0,076 | 0,068 | 0,060 | 0,072 | 0,017 | 0,011 | -0,002 |
| Asset name | PIVX | ETH | XRP | UNO | NEO | DCR | XLM | GRS | XVG | WAVES | |

---

[14] The data referred to the "Traditional portfolio" are provided by investing.com .
[15] The data are provided by cryptomarket.com.
[16] In order to avoid the loss of relevant information and the use of stock prices too far in time, the missing data were treated with the technique of linear interpolation (Noor et al, 2014). The assets that had more than 10% missing data in their series of historical returns are not considered.

| Asset name | PRTP | RWEG | AMA | IUSM | BA | Va | MSFT | PA | HG | GC | CL |
|---|---|---|---|---|---|---|---|---|---|---|---|
| Expected return | 0,66% | 0,38% | 0,55% | 0,53% | 0,54% | 0,48% | 0,52% | 0,72% | 0,79% | 0,37% | |
| Standard Deviation | 0,101 | 0,064 | 0,093 | 0,089 | 0,100 | 0,090 | 0,098 | 0,136 | 0,161 | 0,079 | |
| Kurtosis | 0,841 | 0,297 | 2,551 | -0,615 | 1,518 | 0,869 | 1,750 | 1,943 | 0,989 | 0,286 | |
| Skewness | 3,830 | 3,372 | 27,136 | 20,788 | 11,489 | 2,953 | 11,543 | 10,030 | 7,481 | 2,321 | |
| Adjusted Sharpe | 0,066 | 0,060 | 0,057 | 0,057 | 0,054 | 0,054 | 0,053 | 0,053 | 0,049 | 0,047 | |

A particular GARCH process for the variance is specified for each asset. Table 2 summarized for each asset the best selected ARMA-GARCH models[17], the assumed conditional distribution and the value of the BIC, estimated as indicated in the Methodology section[18].

**Table 2**: **ARMA-GARCH models of cryptocurrencies and financial assets**

| | PRTP | RWEG | AMA | IUSM | BA | Va | MSFT | PA | HG | GC | CL |
|---|---|---|---|---|---|---|---|---|---|---|---|
| E-GARCH model | (2,3) | (1,3) | (2,1) | (2,3) | (1,2) | (2,1) | (3,3) | (1,3) | (2,2) | (3,2) | (3,3) |
| ARMA model | (0,0) | (1,0) | (0,1) | (1,0) | (0,1) | (0,0) | (0,0) | (0,0) | (1,0) | (0,0,0) | (1,0) |
| Cond. distr. | norm | norm | norm | norm | s-norm | norm | norm | norm | norm | norm | s-norm |
| BIC | -5,93 | -6,03 | -6,60 | -8,64 | -6,14 | -6,67 | -6,43 | -6,16 | -6,55 | -7,17 | -5,85 |
| | ETH | XRP | XLM | NEO | DCR | WAVES | GRS | XVG | PIVX | UNO | |
| E-GARCH model | (3,3) | (1,1) | (1,2) | (1,2) | (3,3) | (3,3) | (2,3) | (3,3) | (3,3) | (2,3) | |
| ARMA model | (0,0) | (0,0) | (0,0) | (0,0) | (0,1) | (0,0) | (0,1) | (0,1) | (0,1) | (0,0) | |
| Cond. distr. | norm | s-norm | norm | s-norm | s-norm | s-norm | norm | norm | norm | norm | |
| BIC | -2,75 | -2,44 | -2,15 | -2,08 | -2,11 | -2,35 | -1,57 | -1,37 | -1,97 | -2,21 | |

## 3.1 Estimation of Historical Performance

By using the standardized residuals, it is possible to proceed with the calculus of the correlation among the assets in different ways and to determine the historical performance for each of the three different portfolios above mentioned. The optimization process is carried out minimizing the variance-covariance matrix and maximizing the Sharpe ratio[19] of the portfolio (Bilir, 2016). Table

---

[17] The eGARCH models have always shown the lowest BIC.
[18] The models were estimated through "rugarch" R-package (Ghalanos, 2019).
[19] The Sharpe ratio considered is the original one (Sharpe, 1994).

3 shows the final performances of the three portfolios when the simple "Markowitz optimization" (bounded and not)[20] is applied to the historical returns for each portfolio.

**Table 3**: **Statistical measures and final performance of the simple "Markowitz optimization" approach.**

|  | Trad+Cry | Trad+Cry (bound) | Trad | Trad (bound) | Cryptos | Cryptos (bound) |
|---|---|---|---|---|---|---|
| Expected return | 0,00092 | 0,00090 | 0,00071 | 0,00069 | 0,00584 | 0,00583 |
| Standard Deviation | 0,00741 | 0,00732 | 0,00653 | 0,00644 | 0,06032 | 0,06026 |
| Kurtosis | -0,25233 | -0,26028 | -0,65678 | -0,68337 | -0,28524 | -0,28772 |
| Skewness | 2,18322 | 2,14843 | 4,98413 | 5,01445 | 1,44485 | 1,44738 |
| Adjusted Sharpe | 0,12474 | 0,12318 | 0,10667 | 0,10488 | 0,09702 | 0,09690 |
| Sharpe | 0,12486 | 0,12330 | 0,10895 | 0,10715 | 0,09686 | 0,09674 |

The Copula and Vine Copula approach are introduced in order to unlink the linear correlation from Markowitz optimization. As explained in Section 2, the first step is to transform the standardized residuals into PITs. Table 4 shows the distribution that was individuated for each asset considering the $p$ values of K-S and Uniformity tests.

**Table 4**: **Statistical distributions of standard residuals of cryptocurrencies and financial assets.**

|  | GC | PA | HG | BA | MSFT | Va | RWEG | AMA | PRTP | CL | IUSM |
|---|---|---|---|---|---|---|---|---|---|---|---|
| Distribution | Skwd. GED | Skwd. GED | GED | Skwd. GED | Skwd. GED | Skwd. GED | Skwd. GED | Skwd. GED | Skwd. GED | Skwd. GED | Skwd. GED |
| K - S test | 0,989 | 0,336 | 0,894 | 0,861 | 0,894 | 0,305 | 0,655 | 0,979 | 0,965 | 0,484 | 0,923 |
| PIT test | 0,293 | 0,604 | 0,490 | 0,840 | 0,960 | 0,135 | 0,574 | 0,487 | 0,409 | 0,777 | 0,838 |

|  | ETH | XRP | XLM | NEO | DCR | WAVES | GRS | XVG | PIVX | UNO |
|---|---|---|---|---|---|---|---|---|---|---|
| Distribution | Skwd. GED | Skwd. GED | Skwd. t-Stud. | Skwd. t-Stud. | Skwd. GED | Skwd. GED | Skwd. t-Stud. | Skwd. GED | Skwd. t-Stud. | GED |
| K - S test | 0,998 | 0,785 | 0,989 | 0,947 | 0,965 | 0,655 | 0,998 | 0,824 | 0,979 | 0,158 |
| PIT test | 1,000 | 0,828 | 0,976 | 0,990 | 0,659 | 0,620 | 0,983 | 0,195 | 0,994 | 0,124 |

---

[20] Very often Markowitz optimization tends to allocate the entire weight in a few assets, without considering the principle of diversification (Braga, 2016); to avoid this problem it is possible to set a minimum weight so as to ensure the presence of each asset in the portfolio, at least in a small part. For this empirical analysis a threshold of 1% is set.

The resulting PITs are used as input for the Copula and Vine Copula models. Table 5 shows the *p* values of GoF test, the BIC and the AIC values for the four tested Copula families[21].

**Table 5: Statistical measures of multivariate Copula models for the three portfolios.**

|  | Normal Copula | t Copula | Gumbel Copula | Clayton Copula |
|---|---|---|---|---|
| **Traditonal Portafoglio** | | | | |
| GoF test | 0 | 0,2 | 0 | 0 |
| AIC | -1455,6845 | -1953,7731 | -221,2759 | -427,9208 |
| BIC | -1203,1424 | -1696,6393 | -216,6842 | -423,3291 |
| **Traditonal Crypto Portafoglio** | | | | |
| GoF test | 0 | 0,6 | 0,2 | 0 |
| AIC | -1446,2462 | -1882,4204 | -191,0680 | -405,5577 |
| BIC | -1143,1957 | -1574,7782 | -186,4764 | -400,9660 |
| **Cryptos Portfolio** | | | | |
| GoF test | 0 | 0 | 0 | 0 |
| AIC | -3061,3267 | -3347,5690 | -2539,6445 | -2905,5006 |
| BIC | -2854,7014 | -3136,3520 | -2535,0529 | -2900,9089 |

As it is possible to notice, the t Copula presents the best values for each category of parameters so it will be used to model the correlation structure for all the three portfolios. Table 6 shows the final performances of the three portfolios when the "GARCH-t-copula-Markowitz optimization" (bounded and not) is applied to the historical returns for each portfolio.

---

[21] The Copula models was estimated using "copula" R-package (Hofert et al., 2018; Yan, 2007; Kojadinovic & Yan, 2010; Hofert & Mächler, 2011).

**Table 6**: **Statistical measures and final performance of the "GARCH-t-Copula-Markowitz optimization" approach.**

|  | Trad+Cry | Trad+Cry (bound) | Trad | Trad (bound) | Cryptos | Cryptos (bound) |
|---|---|---|---|---|---|---|
| Expected return | 0,00089 | 0,00087 | 0,00069 | 0,00067 | 0,00612 | 0,00612 |
| Standard Deviation | 0,00715 | 0,00707 | 0,00634 | 0,00625 | 0,06982 | 0,06995 |
| Kurtosis | -0,26356 | -0,27493 | -0,65851 | -0,68937 | 0,20948 | 0,22088 |
| Skewness | 2,22084 | 2,18260 | 4,91895 | 4,94759 | 1,83626 | 1,83579 |
| Adjusted Sharpe | 0,12411 | 0,12258 | 0,10626 | 0,10449 | 0,08828 | 0,08807 |
| Sharpe | 0,12429 | 0,12276 | 0,10849 | 0,10672 | 0,08764 | 0,08742 |

The low flexibility of the traditional Copulas can be solved using the Vine Copula approach. Table 7 shows the correlation structure, the *p* value of GoF test and values of the BIC and AIC for each estimated Vine Copula[22].

**Table 7: Vine Copula structures for "Traditional portfolio", "Traditional Crypto portfolio" and "Cryptos portfolio".**

|  | R Vine | C Vine | D Vine |
|---|---|---|---|
|  | Traditional portfolio | | |
|  | 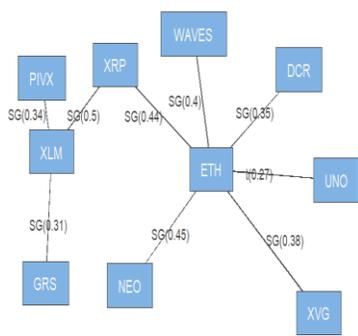 | 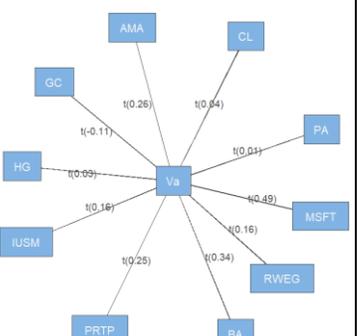 | 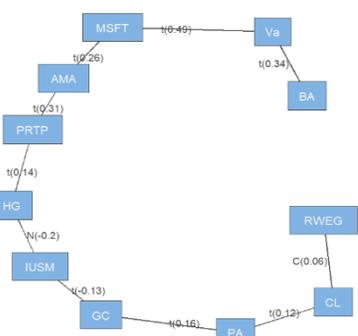 |
| GoF test | 1 | 1 | 1 |
| AIC | -2064,60 | -2038,17 | -2058,40 |
| BIC | -1674,31 | -1611,14 | -1649,74 |
|  | Traditional Crypto portfolio | | |

---

[22] The Vine Copula model with the lowest AIC and BIC values is chosen for each portfolio. The process of identifying the disposition of assets and the Copula families that bind the asset pairs is described by Dissmann et al. (2013). The procedure can be performed in R thanks to the "VineCopula" (Schepsmeier et al., 2018) and "CDVine" (Brechmann & Schepsmeier, 2013) packages.

| | R Vine | C Vine | D Vine |
|---|---|---|---|
| | *(R Vine diagram)* | *(C Vine diagram)* | *(D Vine diagram)* |
| GoF test | 1 | 1 | 1 |
| AIC | -2062,48 | -2048,82 | -2045,47 |
| BIC | -1598,72 | -1571,29 | -1577,12 |

Cryptos portfolio

| | R Vine | C Vine | D Vine |
|---|---|---|---|
| | *(R Vine diagram)* | *(C Vine diagram)* | *(D Vine diagram)* |
| GoF test | 1 | 1 | 1 |
| AIC | -3729,40 | -3737,37 | -3692,75 |
| BIC | -3475,64 | -3476,86 | -3426,44 |

Considering the obtained values, the R-vine copula is used for both "Traditional portfolio" and "Traditional Crypto portfolio"; the C-vine copula is chosen for "Cryptos portfolio". Table 8: shows the final performances of the three portfolios when the "GARCH-Vine Copula Markowitz optimization" (bounded and not) is applied to the historical returns for each portfolio.

**Table 8**: **Statistical measures and final performance of the "GARCH-Vine Copula-Markowitz optimization" approach.**

|  | Trad (bound) | Trad | Trad+Cry | Trad+Cry (bound) | Cryptos (bound) | Cryptos |
|---|---|---|---|---|---|---|
| Expected return | 0,00073 | 0,00056 | 0,00061 | 0,00050 | 0,00463 | 0,00456 |
| Standard Deviation | 0,00740 | 0,00594 | 0,00683 | 0,00593 | 0,05960 | 0,06057 |
| Kurtosis | -0,72972 | -0,59870 | 0,32276 | 0,30873 | -0,36989 | -0,33666 |
| Skewness | 6,16765 | 4,38488 | 2,98568 | 3,71192 | 1,53819 | 1,48117 |
| Adjusted Sharpe | 0,09669 | 0,09218 | 0,08945 | 0,08390 | 0,07759 | 0,07527 |
| Sharpe | 0,09918 | 0,09356 | 0,08902 | 0,08375 | 0,07760 | 0,07523 |

The weights that had been estimated using the returns up to 31 December 2018, have been applied to the future returns that the same assets have in a period between 01 January 2019 and 31 March 2019, in order to evaluate the predictive capacity of the models. Table 9 shows the performances that would have been achieved.

**Table 9**: **Performance and statistics of the three portfolios using the estimated weights at 31 December 2018.**

|  | GARCH t-copula Markow approach | | | | | |
|---|---|---|---|---|---|---|
|  | Trad (bound) | Trad | Trad+Crypto (bound) | Trad+Crypto | Cryptos | Cryptos (bound) |
| Expected return | 0,00167 | 0,00166 | 0,00165 | 0,00164 | 0,00232 | 0,00228 |
| Standard Deviation | 0,00703 | 0,00712 | 0,00713 | 0,00721 | 0,04865 | 0,04832 |
| Kurtosis | 0,37768 | 0,35138 | 0,39594 | 0,37523 | 1,24711 | 1,26607 |
| Skewness | 1,50712 | 1,44536 | 1,52881 | 1,42619 | 6,82703 | 6,90927 |
| Adjusted Sharpe | 0,24525 | 0,24060 | 0,23871 | 0,23445 | 0,04790 | 0,04724 |
| Sharpe | 0,23815 | 0,23386 | 0,23187 | 0,22781 | 0,04779 | 0,04713 |

|  | GARCH Markow approach | | | | | |
| --- | --- | --- | --- | --- | --- | --- |
|  | Trad (bound) | Trad | Trad+Crypto (bound) | Trad+Crypto | Cryptos | Cryptos (bound) |
| Expected return | 0,00170 | 0,00169 | 0,00167 | 0,00166 | 0,00106 | 0,00104 |
| Standard Deviation | 0,00728 | 0,00738 | 0,00751 | 0,00760 | 0,03711 | 0,03690 |
| Kurtosis | 0,48435 | 0,45807 | 0,47354 | 0,45277 | -0,41620 | -0,41662 |
| Skewness | 1,48944 | 1,44603 | 1,51607 | 1,44284 | 1,76452 | 1,78371 |
| Adjusted Sharpe | 0,24141 | 0,23658 | 0,22997 | 0,22569 | 0,02867 | 0,02811 |
| Sharpe | 0,23357 | 0,22917 | 0,22297 | 0,21896 | 0,02869 | 0,02813 |
|  | GARCH-Vine copula-Markow approach | | | | | |
|  | Trad (bound) | Trad | Trad+Crypto (bound) | Trad+Crypto | Cryptos (bound) | Cryptos |
| Expected return | 0,00195 | 0,00155 | 0,00102 | 0,00102 | 0,00087 | 0,00085 |
| Standard Deviation | 0,00783 | 0,00639 | 0,00574 | 0,00644 | 0,03219 | 0,03251 |
| Kurtosis | 0,72057 | 0,08311 | 0,32236 | 0,06806 | -0,95559 | -0,91788 |
| Skewness | 2,34643 | 2,01057 | 0,28685 | 0,61119 | 4,83958 | 4,88716 |
| Adjusted Sharpe | 0,25813 | 0,24592 | 0,18235 | 0,16066 | 0,02691 | 0,02592 |
| Sharpe | 0,24899 | 0,24268 | 0,17712 | 0,15789 | 0,02708 | 0,02608 |

### 3.2 Forecasting and possible evolution of portfolio

It is possible to make forecasting by using the estimated models. In the Copulas model the random variables extraction from the Copula structure is used for forecasting; in the Vine Copula approach the simulation algorithm used to make predictions is described by Dissmann (Dissmann et al., 2013).[23] The empirical analysis considers two points of view: the first one assesses whether the method of calculation of the correlation structure should change considering different investment time horizons (Table 11)[24]; the second one investigates on the stability of the performance associate to each asset considering a rolling investment window: at the end of each time window within the forecast period, the portfolio is rebalanced. The forecasting period is considered to be 90 days (Table 10).[25]

---

[23] Using the properties of the Location-Scale model, once the standardised residues have been simulated (for each of the two procedures), it is possible to multiply them by the conditioned standard deviation and add the conditioned

**Table 10**: **Performances and statistics of three portfolios considering a 30-days rebalancing.**

| | First 30 days | | | | | |
|---|---|---|---|---|---|---|
| | Trad cop. | Trad cop. vine | Trad+Crypto cop vine | Trad+Crypto cop. | Cryptos cop. vine | Cryptos cop. |
| Expected return | 0,00196 | 0,00192 | 0,00183 | 0,00178 | -0,00363 | -0,00381 |
| Standard Dev. | 0,00599 | 0,00605 | 0,00644 | 0,00629 | 0,03389 | 0,03401 |
| Kurtosis | 0,36503 | 0,32307 | 0,29744 | 0,33120 | -0,69873 | -0,63477 |
| Skewness | 0,07012 | 0,15761 | 0,28638 | 0,19303 | 0,99652 | 0,92418 |
| Adjusted Sharpe | 0,34730 | 0,33507 | 0,29795 | 0,29732 | -0,10740 | -0,11232 |
| Sharpe | 0,32766 | 0,31768 | 0,28476 | 0,28349 | -0,10702 | -0,11208 |
| | Second 30 days | | | | | |
| | Trad cop. | Trad cop.vine | Trad+Crypto cop. vine | Trad+Crypto cop. | Cryptos cop. | Cryptos cop. vine |
| Expected return | 0,00186 | 0,00185 | 0,00186 | 0,00184 | 0,00355 | 0,00349 |
| Standard Dev. | 0,00392 | 0,00392 | 0,00397 | 0,00399 | 0,03210 | 0,03217 |

---

average (where the future estimates of both conditioned measures are calculated using the ARMA-GARCH model selected earlier), obtaining an estimate of the returns for each day of the forecast period. 5000 simulations of a matrix are carried out: it has as columns the number of assets that make up the portfolio and as rows the number of days for which the forecast is made.

[24] Three investment time horizons of 30, 60 and 90 days are considered, for each of which a process of Markowitz optimization takes place.

[25] The time window is fixed at 30 days and therefore the portfolio will be rebalanced and optimized 3 times within the forecast period.

|  | | | | | | |
|---|---|---|---|---|---|---|
| Kurtosis | -0,18840 | -0,18363 | -0,23582 | -0,29604 | -1,25406 | -1,23559 |
| Skewness | -0,28078 | -0,28845 | -0,52100 | -0,59023 | 3,92928 | 3,93804 |
| Adjusted Sharpe | 0,49734 | 0,49551 | 0,49235 | 0,48299 | 0,10745 | 0,10563 |
| Sharpe | 0,47371 | 0,47182 | 0,46875 | 0,46163 | 0,11048 | 0,10851 |

| | Third 30 days | | | | | |
|---|---|---|---|---|---|---|
| | Cryptos cop. vine | Cryptos cop. | Trad+Crypto cop. vine | Trad+Crypto cop. | Trad cop. vine | Trad cop. |
| Expected return | 0,00416 | 0,00405 | 0,00050 | 0,00045 | 0,00034 | 0,00034 |
| Standard Dev. | 0,02482 | 0,02453 | 0,00417 | 0,00421 | 0,00435 | 0,00438 |
| Kurtosis | -0,10121 | -0,15886 | -0,22111 | -0,34644 | -0,20882 | -0,29315 |
| Skewness | 0,83759 | 0,89118 | 1,10909 | 1,19796 | 1,31970 | 1,37473 |
| Adjusted Sharpe | 0,16972 | 0,16661 | 0,11948 | 0,10775 | 0,07860 | 0,07667 |
| Sharpe | 0,16766 | 0,16494 | 0,11889 | 0,10755 | 0,07839 | 0,07656 |

**Table 11: Performances and statistics of three portfolios considering three different time horizons.**

| | Time horizon: 30 day | | | | | |
|---|---|---|---|---|---|---|
| | Trad cop. | Trad cop.vine | Trad+Crypto cop. vine | Trad-Crypto cop. | Cryptos cop. vine | Cryptos cop. |
| Expected return | 0,00196 | 0,00192 | 0,00183 | 0,00178 | -0,00363 | -0,00381 |
| Standard Dev. | 0,00599 | 0,00605 | 0,00644 | 0,00629 | 0,03389 | 0,03401 |
| Kurtosis | 0,36503 | 0,32307 | 0,29744 | 0,33120 | -0,69873 | -0,63477 |
| Skewness | 0,07012 | 0,15761 | 0,28638 | 0,19303 | 0,99652 | 0,92418 |
| Adjusted Sharpe | 0,34730 | 0,33507 | 0,29795 | 0,29732 | -0,10740 | -0,11232 |
| Sharpe | 0,32766 | 0,31768 | 0,28476 | 0,28349 | -0,10702 | -0,11208 |

|  | Time horizon: 60 day | | | | | |
|---|---|---|---|---|---|---|
|  | Trad cop. | Trad cop. vine | Trad+Crypto cop. vine | Trad+Crypto cop. | Cryptos cop. vine | Cryptos cop. |
| Expected return | 0,00202 | 0,00199 | 0,00196 | 0,00192 | -0,00004 | -0,00006 |
| Standard Dev. | 0,00547 | 0,00550 | 0,00568 | 0,00565 | 0,03362 | 0,03365 |
| Kurtosis | 0,29801 | 0,25694 | 0,24267 | 0,24159 | -0,91550 | -0,91493 |
| Skewness | 0,85303 | 0,95238 | 1,08142 | 0,98918 | 2,30137 | 2,28767 |
| Adjusted Sharpe | 0,38744 | 0,37800 | 0,35984 | 0,35360 | -0,00121 | -0,00188 |
| Sharpe | 0,36854 | 0,36127 | 0,34547 | 0,33932 | -0,00121 | -0,00188 |

|  | Time horizon: 90 day | | | | | |
|---|---|---|---|---|---|---|
|  | Trad cop. vine | Trad cop. | Trad+Crypto cop. | Trad+Crypto cop. vine | Cryptos cop. vine | Cryptos cop. |
| Expected return | 0,00147 | 0,00146 | 0,00145 | 0,00146 | 0,00129 | 0,00126 |
| Standard Dev. | 0,00569 | 0,00572 | 0,00574 | 0,00579 | 0,03103 | 0,03112 |
| Kurtosis | 0,13025 | 0,09857 | 0,12489 | 0,13367 | -0,85149 | -0,84144 |
| Skewness | 1,34281 | 1,38368 | 1,42199 | 1,46765 | 2,83125 | 2,71640 |
| Adjusted Sharpe | 0,26408 | 0,26030 | 0,25869 | 0,25762 | 0,04146 | 0,04033 |
| Sharpe | 0,25804 | 0,25485 | 0,25314 | 0,25214 | 0,04170 | 0,04054 |

## 4. Discussion of Results

Referring to the results just shown, some considerations are mandatory. The results in table 1 show that resulting in an averagely higher Adjusted Sharpe Ratio, cryptocurrencies, are not excessively riskier as compare to traditional assets. While the results in table 2 confirm the presence of a *leverage effect* (Bouchaud et al, 2001), which is also the example of a generalized goodness of fit of the EGARCH model. The results in table 3 and 6 show that the best performance is achieved by the "Traditional Crypto portfolio" with the "Markowitz optimization" approach and for "Traditional Crypto portfolio" through the "GARCH-t Copula Markowitz optimization" approach. While, results of table 8 show that the "GARCH-Vine Copula Markowitz optimization" approach, on the other hand, resulted in worst performance. Results in table 9, suggest that with allocation of weights to returns of assets through "GARCH-Vine Copula Markowitz optimization" the "Traditional Portfolio" outperform other portfolios. This result offers the possibility to use approaches with a statistical structure (GARCH-Cupula and GARCH-Vine Copula) in order to perform simulations and to use them to make forecasts.

Table 10 shows the result obtained in the case of periodic rebalancing of the portfolio: a careful analysis of the two central columns validate that, regardless of the statistical process underlying the simulation, the "Traditional Crypto portfolio" performances maintain it in second position, without ever having an excessive gap from the best performing portfolio in a certain sub-period, while, the other portfolios show drastic changes in performance[26], the "Traditional Crypto portfolio" results stable relatively to each considered period (further confirmation of the diversification effect of the cryptocurrencies).

Results of table 11 confirms that, the temporal horizon of investment results, not to be a conditioning variable when one of the two proposed methods (Copula and Vine Copula) is applied to calculate the correlation between the assets. In fact, considering the 3 portfolios coupled for simulative method (Copula and Vine Copula), it is possible to see as the reached performances do not vary in a substantial way: therefore, there is no more reason to assert that it should be used as a method, according to the change of the investment horizon.

---

[26] Note that the "Crypto portfolio" had been the least performing during the first 30 days, while after the third rebalancing it showed the best results; vice versa for the "Traditional portfolio".

Moreover, table 3, 5, 6 and 8 reflect that a portfolio composed exclusively of cryptocurrencies has a lower performance as compare to a "mixed" or a"traditional" portfolio. The results in these tables thus highlight in identifying the correlation structure of cryptocurrencies[27], with the consequence of under- fitting of the models used, due to a market which is not structurally mature and hard to predict.

## 5. Conclusion and Future research directions

The purpose of this analysis is to evaluate the performance of cryptocurrencies as financial assets. To this end, we have constructed three portfolios: the first one consists of traditional assets. We conducted the analysis using autoregressive models to estimate the evolutionary processes of the average and variance of returns. Additionally, we employed Copula and Vine Copula models to assess the correlation structure of the assets and conduct simulations, aiming to investigate the future behavior of the portfolios.

Our findings indicate that, from a post-performance evaluation perspective, the introduction of cryptocurrency into the portfolio yields better results than a portfolio comprised solely of equities, bonds, and commodities or one consisting exclusively of cryptocurrencies. When analyzing the future evolution of the portfolio, the 'Traditional Crypto portfolio' emerges as the most stable in terms of performance. This stability is accompanied by strong performance compared to the best-performing portfolios during certain periods.

Furthermore, we observed that the Copula and Vine Copula approaches exhibit similar simulative capabilities. The differences in representing the correlation structure of assets between these two models become apparent when evaluating past performance.

Empirical evidence suggests that a cryptocurrency portfolio exhibits a lower generalized performance.

This article's concept can be further expanded by considering not only evolutionary processes for the average and variance but also for correlations, introducing dynamic Copula models (So and

---

[27] The single cryptocurrency has generally better performance than traditional assets (Table 1).

Yeung, 2014; Aepli et al., 2015). Future studies should focus on measuring risk and optimizing portfolio performance when risk is not solely expressed by a single measure. For this purpose, a polynomial goal programming (PGP) model could be proposed (Aracioğlu, Demircan, and Soyuer, 2011; Lai, Yu, and Wang, 2006).

In addition to methodological considerations, further research should explore how many and which cryptocurrencies are needed to create an optimal portfolio (Songa, Changa, and Songd, 2019; Antonakakis, Chatziantoniou, and Gabauer, 2019).

Lastly, we recommend applying portfolio optimization for investments with cryptoassets in the midterm and long term based on the Markowitz approach, especially as long as the cryptomarkets remain unregulated and blockchain technology is not widely adopted. This recommendation is based on economic considerations rather than solely on statistical facts.